\documentclass{gen-j-l}

\usepackage{amssymb}

\newtheorem{theorem}{Theorem}[section]

\theoremstyle{definition}

\theoremstyle{remark}

\numberwithin{equation}{section}

\providecommand{\R}{{\mathbb R}}
\newtheorem{algorithm}{Algorithm}[section]

\begin{document}

\title{Nonsingular Efficient Modeling of \\ Rotations in 3-space using
three components}


\author{Norman J. Goldstein }

\address{Research \& Development, MacDonald Dettwiler and Associates \\
13800 Commerce Parkway, Richmond, BC, Canada}
\curraddr{}
\email{norman@mdacorporation.com}
\thanks{The work by which this paper came to fruition was partly funded by
Defense Research and Development Canada (DRDC).  I wish also to 
acknowledge the support of the R\&D department at MacDonald Dettwiler
and Associates, especially Vinay Kotamraju and Pezhman Firoozfam for
comments and suggestions.}



\date{}

\dedicatory{}


\begin{abstract}
This article introduces yet another representation of
rotations in 3-space.  The rotations form a 3-dimensional
projective space, which fact has not been exploited in 
Computer Science.  We use the four affine patches of this
projective space to parametrize the rotations.  This affine
patch representation is more compact than quaternions (which
require 4 components for calculations), encompasses the entire
rotation group without singularities 
(unlike the Euler angles and rotation vector approaches), and requires only 
ratios of linear or quadratic polynomials for basic computations 
(unlike the Euler angles and rotation vector approaches 
which require transcendental functions).

As an example, we derive the differential equation
for the integration of angular velocity using this affine patch
representation of rotations.  We remark that the complexity of
this equation is the same as the corresponding quaternion equation,
but has advantages over the quaternion approach e.g.
renormalization to unit length is not required, and state space has
no ``dead'' directions.
\end{abstract}

\maketitle


\section{Introduction}

This work was motivated by the need to have a practical and robust way to
model rotations and projective spaces in the software that we write. These
are applications in photogrammetry, remote sensing and simultaneous
localization and mapping.  Rotations are used in modeling the poses of the
cameras, while the focal planes of the cameras are projective spaces.  
Although it is true that the imaging focal plane is contained in a single 
affine patch 
of ${\R}P^2$, relevant vanishing points may lie on, or near, the line at 
infinity, so that the coordinates of the imaging affine patch are not 
appropriate for robustly describing and managing the vanishing points.

Traditionally, (unit) quaternions have been used to model rotations,
as the quaternions have only 4 components -- just one more parameter
than the dimension of the space of rotations (see, for example
\cite{schultz06, rogers03}).  However, the extra component is
bothersome, requiring extra storage, requiring extra work to manage
the redundancy (renormalization to unit length), and likely affecting
the convergence rate due to motion in the redundant {\em length}
direction of the quaternion, when we are optimizing to find the best
rotation for a particular problem.

Another approach to representing
rotations is to use the three {\em Eulerian angles}, e.g.\cite{goldst50}.
However, this approach suffers from not being everywhere invertible, so
is not a global solution to representing the rotations.  Moreover, the
transcendental transformation between Eulerian angles and rotation 
matrices is not ideal for software implementation.  A similar situation
is to be found with the {\em rotation vector} representation of a rotation.
The on-line article \cite{wiki-rot} discusses this and other representations
of rotations.

In the following sections, we review the standard coordinates on
projective space, and then derive the differential equation for the
integration of angular velocity using the projective space manifold
structure (affine patches) on the rotation group.  
We will see that with this novel representation of the rotations,
the differential equation has the same
complexity as the corresponding quaternion differential equation, and
has other desirable properties over the quaternion case.

\section{Notation}
All vector spaces and projective spaces are over the real numbers.
For a point $z$ in n-dimensional projective space, $P^n$, its homogeneous
coordinates are denoted $z = [z^0, z^1, \ldots z^n]$.

The traditional manifold structure
on $P^n$ is given as

\begin{itemize}
\item The $k^{th}$ affine patch, $k = 0, \ldots n$,  on $P^n$ is defined to be the set
\begin{equation*}
  V_k = \{ z = [z^0, z^1, \ldots z^n] : z^k \neq 0 \}
\end{equation*}
\item The coordinate chart for $V_k$ is 
\begin{equation*}
x_k : V_k \longrightarrow \R^n
\end{equation*}
defined by the formula $x_k(z) = z / z^k$ and throw away the 1 in the $k^{th}$ slot.
This is seen to be a 1-1 map onto all of $\R^n$.
\end{itemize}

The following algorithm template illustrates how to use the 
manifold structure on $P^n$ to iteratively step along a sequence of points
in projective n-space.  For simplicity, we do not specify what it means
to ``iterate'', and we omit any stopping criteria.

\begin{algorithm}{Iterate on Projective Space} 
\begin{enumerate}
\item[ Input ] An initial point, $z_0 \in P^n$.
\item Let k be the index of the largest component, $|{z_0}^k|$.
\item Represent $z_0$ in the $k^{th}$ affine patch.  So, initially,
all the coordinate components in this patch have magnitude at most 1.
\item Iterate to the next value of $z$, in this patch.
\item If all coordinate components of $z$ have magnitude at most 2, 
go back to Step 3.  Otherwise, proceed to Step 5.
\item Reconstruct the homogeneous coordinates, z, for the current
point, and let k be the index of the the largest component $|z^k|$.
\item Represent $z$ in the $k^{th}$ affine patch, and proceed
to Step 3.
\end{enumerate}
In Step 6, some book-keeping may be required to keep the overall
environment of the algorithm consistent when switching patches.  The
choice of the value, 2, in Step 4 is somewhat arbitrary, but has worked
well for us.  The point is to monitor when a particular patch is no longer
appropriate for the current value of $z$, and then to switch to a good patch.
If $z$ begins to converge, the switching of patches will cease.  
\end{algorithm}
This algorithm may be used when optimizing for a ``best'' value of z,
or when incrementally solving a differential equation on projective space.
In the next section, we write down explicitly the differential equation for
integrating angular velocity, using the projective space manifold structure
of the rotations.

\section{Integrating Angular Velocity}

We use the convention that the components of a quaternion, $q$, are
indexed as $q^0, q^1, q^2$ and $q^3$, with $s = q^0$ being the scalar
component, and $v$ denoting the 3-vector of the other (spatial) components.  

The particular rotation matrix associated to q is
\begin{equation*}
R = I (2 s^2 - 1) + 2vv^t + 2sM_v
\end{equation*}

Where, $I$ is the identity 3x3 matrix, and  
$M_a$ is the usual 3x3 skew matrix defined by 
\[ M_a b = a \times b \]
Explicitly,
\begin{equation*}
M_a = \left(
\begin{array}{ccc}
0       & -a^3   & a^2 \\
a^3   &   0      & -a^1 \\
-a^2  &   a^1  &   0
\end{array}
\right)
\end{equation*}

Our jumping off point for analyzing angular velocity is the well-known
(\cite{schultz06}\cite{rogers03})
differential equation relating angular velocity, $\omega$, and quaternions:
\begin{equation}
\label{equ:dqdt}
\frac{dq}{dt} = \frac{1}{2} H q = \frac{1}{2} 
\left(
\begin{array}{cc}
0 &\ \  -\omega^t \\
\omega & \ \ -M_\omega
\end{array}
\right)
q
\end{equation}

This corresponds to the matrix equation
\begin{equation*}
\frac{dR}{dt} = R M_\omega \ .
\end{equation*}
 
It is Equation \ref{equ:dqdt} that we wish to translate to the four affine 
patches of the rotations.  To that end, we make explicit the components 
of the matrix in Equation~\ref{equ:dqdt}.
\begin{equation}
H = \left(
\begin{array}{cccc}
0                 &     -\omega^1   &   -\omega^2  &    -\omega^3   \\
\omega^1  &          0             & \omega^3    & -\omega^2         \\
\omega^2  &  -\omega^3    &        0            & \omega^1          \\
\omega^3  &   \omega^2   &   -\omega^1  &        0
\end{array}
\right)
\end{equation}

We restrict attention to patch $i$, $i = 0, 1, 2, 3$, and let
\[
x = \left(
\begin{array}{c}
x^1 \\
x^2 \\
x^3
\end{array}
\right)
\]
be the 3 components parametrizing the rotations from this patch.  The corresponding
homogeneous coordinates, $\hat{x}$ are obtained by forming the 4-vector with a 1
in the $i^{th}$ slot.  So, $\hat{x}$ is a non-unit quaternion, and we define a time-varying
scale factor, $\alpha = \alpha(t) $ by the equation
\begin{equation*}
 \hat{x} = \alpha q
\end{equation*}
where q is now a unit quaternion.  Consequently, Equation \ref{equ:dqdt} 
is valid for q, and we may write

\begin{equation*}
\frac{d \hat{x}}{dt} = \frac{d \alpha}{dt} q + \alpha \frac{1}{2} H q
\end{equation*}
which may be rewritten as
\begin{equation}
\label{equ:dxhatdt}
\frac{d \hat{x}}{dt} = \beta \hat{x} + \frac{1}{2} H \hat{x}
\end{equation}
where $ \beta = \frac{d \alpha}{dt} / \alpha $.  To reduce this equation, further,
we define the 3x4 matrix whose columns are the non-zero entries from the
corresponding columns of H.

\begin{equation}
W = \left(
\begin{array}{cccc}
\omega^1  &    -\omega^1   &   -\omega^2  &    -\omega^3   \\
\omega^2  &  -\omega^3    & \omega^3    & -\omega^2         \\
\omega^3  &   \omega^2   &   -\omega^1  & \omega^1      
\end{array}
\right)
\end{equation}
We recall that the $i^{th}$ component of $\hat{x}$ is identically 1, so its derivative is 0,
and we conclude from the $i^{th}$ component of Equation \ref{equ:dxhatdt} that 
\begin{equation*}
0 = \beta + \frac{1}{2} H^i \hat{x}
\end{equation*}
where $H^i$ is the $i^{th}$ row of $H$.  This can be rewritten as
\begin{equation}
\label{equ:beta}
\beta = \frac{1}{2} W_i \cdot x
\end{equation}
where $W_i$ is the $i^{th}$ column of W.

Extracting, now, the non-$i^{th}$ components of Equation \ref{equ:dxhatdt}, we obtain
\begin{eqnarray}
\label{equ:dxdt}
\frac{dx}{dt} & = & \beta x + \frac{1}{2} \left(  W_i +  H^{(i)} x  \right)
\end{eqnarray}
where $H^{(i)}$ is the 3x3 matrix obtained from $H$ by deleting the $i^{th}$ row
and the $i^{th}$ column.  By inspecting the four cases, we see that
\begin{equation}
\label{equ:Hi}
H^{(i)} = (-1)^{i+1} M_{W_i}
\end{equation}
Inserting Equations \ref{equ:beta} and \ref{equ:Hi} into Equation \ref{equ:dxdt}, we
obtain
\begin{eqnarray}
 \frac{dx}{dt} & = & 
\frac{1}{2} (W_i \cdot x) x + \frac{1}{2} \left(  W_i +   (-1)^{i+1} W_i \times x \right)  \nonumber \\
                     & = & \frac{1}{2} \left[  W_i +  (W_i \cdot x) x + (-1)^{i+1} W_i \times x \right] 
\label{equ:final}
\end{eqnarray}

Equation \ref{equ:final} is the 3-dimensional differential equation to integrate angular
velocity using affine patches on the rotation group.
 The RHS contains 12 multiplications, excluding the scaling by 1/2.  This is
the same number of multiplications as in Equation \ref{equ:dqdt}, which is the 
corresponding quaternion differential equation.

\subsection{Difference Equation}

In practice, a difference equation is used to integrate
angular velocity.  This section shows how to obtain
identical results to the quaternion solution, but by using
the affine patch representation of the rotations.  This has 
the advantage that explicit renormalization to unit length is not required,
as it is inherent to the affine patch approach.

For a time step, of length $\Delta t$, we wish to find
the $x$ increment, $\Delta x$, so that the two expressions,
$q + \frac{1}{2}Hq\Delta t$ and $x + \Delta x$,
represent the same rotation i.e.

\begin{equation*}
\frac{ (q + \frac{1}{2}Hq\Delta t) ^{\bar i} }
     { (q + \frac{1}{2}Hq\Delta t) ^i  } = 
x + \Delta x
\end{equation*}

where $\bar{i}$ denotes all the components except for the $i^{th}$.
Divide the top and bottom of the LHS by $q^i$ to obtain

\begin{eqnarray}
\frac{ x + \frac{1}{2}H^{\bar i} \hat{x}\Delta t }
     { 1 + \frac{1}{2}H^i \hat{x} \Delta t } & = & 
x + \frac{\Delta x}{\Delta t} \Delta t
\label{eqn:diff0}
\end{eqnarray}

We recall from the derivation of Equation \ref{equ:final} that
\begin{eqnarray*}
H^{\bar i} \hat{x} & = & W_i - (-1)^i W_i \times x \\
H^i \hat{x} & = & - W_i \cdot x 
\end{eqnarray*}

so that Equation \ref{eqn:diff0} simplifies to
\begin{equation}
\frac{\Delta x}{\Delta t} =
\frac{W_i + (-1)^{i+1}W_i \times x + (W_i \cdot x) x}
     {2 - (W_i \cdot x) \Delta t}
\label{eqn:diff}
\end{equation}

It is interesting to note that this value of $\Delta x/\Delta t$
is the same as dividing $dx/dt$ by the scale factor
$1 - (W_i \cdot x) \Delta t / 2$.

\section{Conclusions}
\label{sec:conclusion}

We have applied elementary manifold theory to construct a global, efficient,
singularity-free parametrization for the rotation group.  We have
derived differential and difference equations 
for integrating angular velocity
using this novel representation of the rotations, and have shown that
the complexity of using these equations is on a par with
the corresponding quaternion equations.  On a theoretical
basis, however, the difference equation
derived in this article is
more sound, as there are no renormalizations to unit
length required.

\bibliographystyle{amsplain}

\end{document}